\documentclass[12pt]{iopart}

\usepackage{graphicx}		

\begin{document}

\title[Mobility of Min-proteins]{Mobility of Min-proteins in \textit{Escherichia coli} measured by fluorescence
correlation spectroscopy}

\author{G Meacci$^{1,*}$, J Ries$^{2,*}$, E Fischer-Friedrich$^{1}$, N 
Kahya$^{2}$, P Schwille$^{2}$ and K Kruse$^{1}$\footnote[3]{Present address: Theoretische Physik, 
Universit\"at des Saarlandes, Postfach 151150, 66041 Saarbr\"ucken, Germany. \ead{karsten@lusi.uni-sb.de}}}

\address{$^1$ Max-Planck Institute for the Physics of Complex Systems,
N\"othnitzerstrasse 38, 01187 Dresden, Germany}
\address{$^2$ TU-Dresden, Biotec, Am Tatzberg 47-51, 01307 Dresden, Germany}
\address{$^{*}$These authors contributed equally to the work.}

\begin{abstract}
In the bacterium \textit{Escherichia coli}, selection of the division site involves 
pole-to-pole oscillations of the proteins MinD and MinE. Different oscillation mechanisms 
based on cooperative effects between Min-proteins and on the exchange of Min-proteins 
between the cytoplasm and the cytoplasmic membrane have been proposed. 
The parameters characterizing the dynamics of the Min-proteins \textit{in vivo} are not known. 
It has therefore been difficult to compare the models quantitatively with experiments.
Here, we present \textit{in vivo} measurements of the mobility of MinD 
and MinE using fluorescence correlation spectroscopy. Two distinct time-scales are
clearly visible in the correlation curves. While the faster time-scale can be attributed
to cytoplasmic diffusion, the slower time-scale could result from diffusion of 
membrane-bound proteins or from protein exchange between the cytoplasm and the
membrane. We determine the diffusion constant of cytoplasmic MinD to be approximately 
$16\mu{\rm m}^2/s$, while for MinE we find about $10\mu{\rm m}^2/s$, 
independently of the processes responsible for the slower time-scale. Implications of the 
measured values for the oscillation mechanism are discussed.
\end{abstract}

\pacs{87.16.-b,87.15.Kg,87.15.Rn}

\maketitle

\section{Introduction}

The rod-shaped bacterium \textit{Escherichia coli} usually divides in 
the center of its long axis between the two segregated copies of its 
chromosome. The position of the division plane is determined by the 
Z-ring, a structure built from the protein FtsZ, which is associated with 
the cytoplasmic membrane and encircles the cytoplasm~\cite{lutk02}. 
Assembly of the Z-ring is targeted to the cell center by two mechansims. 
For one, formation of the ring around the two copies of the chromosome is 
inhibited by proteins binding to DNA, a mechanism termed
nucleoid occlusion~\cite{yu99,bern05}. For the other, the proteins MinC, MinD,
and MinE suppress ring formation close to the cell poles~\cite{boer89,bi93}, 
leaving the center as the only possible site. MinC is able to 
depolymerize FtsZ filaments, while MinD and MinE direct MinC to the cell poles. 
The spatial distributions of MinD and MinE, and hence of MinC, periodically 
change in time: after dwelling in the vicinity of one pole for about 40s, the 
proteins get redistributed to the opposite pole~\cite{rask99,hu99}. These 
oscillations do not require the presence of MinC.

Theoretical works have provided strong evidence that the pole-to-pole oscillations
are formed by self-organization of MinD and MinE~\cite{howa05}. All mechanisms
proposed so far rely essentially in one way or another on the formation of aggregates 
of membrane-bound MinD. Such aggregates have been observed \textit{in vitro} and 
\textit{in 
vivo}~\cite{hu02,shih03}. The mechanisms can roughly be divided into two classes.
In cooperative attachment (CA) models, MinD-aggregates are formed through 
collective effects during binding to the cytoplasmic
membrane~\cite{mein01,howa01,huan03,drew05,pavi06}. In aggregation current (AC)
models, aggregates are formed by mutual attraction after the proteins have bound to the 
membrane~\cite{krus02,meac05}. CA as well as AC models can capture the qualitative 
features of the Min-oscillations and there is experimental evidence for both processes 
in \textit{E.~coli}. The strongest hint for aggregation currents is provided by a study of 
MinD attachment to phospholipid vesicles 
in the presence of ATP$\gamma$S, a non-hydrolyzable ATP analog~\cite{hu02}.  
This work suggests a 
two-step mechanism for the formation of aggregates of membrane-bound MinD 
involving first the binding of MinD to the membrane and subsequent 
aggregation. Furthermore,  using a yeast two-hybrid assay MinD-MinD interactions were 
shown to be stronger if both proteins were membrane-bound than if at least 
one partner was cytoplasmic~\cite{tagh06}. On 
the other hand, the 
concentration-dependence of MinD binding to phospholipid membranes deviates
from Langmuir isotherms~\cite{lack03}. In addition, the amount of MinD binding 
to liposomes
as a function of the MinD-concentration in the surrounding solution
could be fitted by a Hill equation with a
Hill coefficient of 2~\cite{mile03}. These two findings clearly suggest some cooperativity 
during MinD binding to the membrane.

In order to reveal whether cooperative attachment or an aggregation current is 
the primary cause of the Min oscillations, a quantitative comparison of the models 
with experiments is necessary. This requires, in particular, to fix the model parameters 
by measurements. There are several techniques to measure protein mobilities using 
fluorescence microscopy. Direct measurements of the displacement of 
individual proteins have been used to determine the mobility of membrane proteins 
in \textit{Caulobacter crescentus}~\cite{deic04}. Fluoresence recovery after 
photobleaching (FRAP), where the fluorescent proteins present in a defined region 
are bleached and the recovery of the fluorescence is monitored, was used to measure 
the diffusion constants of cytoplasmic proteins~\cite{elow99}. Fluorescence correlation 
spectroscopy (FCS) exploits the fluctuations in the fluorescence intensity emanating 
from an illuminated region in order to assess dynamic 
properties~\cite{kric02}. To this end, the autocorrelation function is measured for 
fluctuations around the mean signal. Fitting this to the autocorrelation curve theoretically 
expected for the process under study then yields the searched-for values.
In bacteria, FCS was used to measure the concentration of phosphorylated CheY 
involved in chemotaxis~\cite{cluz00} and transcription activity at the RNA 
level~\cite{le05,le06}.

We have used FCS to measure the mobility of MinD and MinE tagged to Green 
Fluorescent Protein (GFP) in \textit{E. coli}. We found that a simple diffusion process 
cannot account for the measured autocorrelation curves. We have analyzed the
data assuming that either the mobility of membrane-bound proteins or the binding-unbinding
dynamics
is dominant, and thus obtained key parameters of the various models.
As a control we also measured the mobility of GFP  and found significant deviations 
to previous measurements ~\cite{elow99}.

\section{Materials and Methods}

\subsection{Strains}
EGFP and His6-EGFP were expressed in BL21(DE3)pLysS using the vectors
pBAT4 and pET9d, respectively (Novagen, CN Biosciences).
GFP-MinD was expressed in JS964~\cite{hu03} (J. Lutkenhaus, U. Kansas, USA) and 
WM1255~\cite{corb02} (W. Margolin, U. Texas, USA), and MinE-GFP in 
WM1079~\cite{corb02} (W. Margolin, U. Texas, USA).
Bacteria were grown overnight in 3ml LB medium at 37$^{\circ}$C together with
a concentration of 25$\mu g/ml$ Spectinomycin, 25$\mu g/ml$ Kanamycin,
20$\mu g/ml$ Chloramphenicol and 50$\mu g/ml$ Ampicillin. Of the overnight 
culture 500$\mu l$ were put in 50$ml$ of fresh LB medium containing the same 
concentration of antibiotics as above, and grown at 37$^{\circ}$C until the optical 
density (OD) at 600$nm$ reached $\approx$0.2. Expression of GFP-MinD
in JS964 and His6-EGFP in BL21(DE3)pLysS was induced by adding 20$\mu M$ 
isopropyl-$\beta$-D-thiogalactopyranoside (IPTG). Expression of MinE-GFP in WM
1079 was induced by adding 0.005$\%$ L-arabinose. No inducer was used for 
GFP-MinD expression in WM1255. Then the bacteria were grown at 30$^{\circ}$C 
for 1-2 hours, usually sufficient to produce visible fluorescence and to see Min  
oscillations. To reduce background fluorescence from the buffer, LB medium was 
prepared with 1g of yeast extract per liter. For microscopy an approximatelty $0.5$mm thick 
solid slab of $1\%$ agarose (Invitrogene, 15510-027) in LB medium was prepared 
between a 25mm$\times$75mm glass slide and a 18mm$\times$18mm cover slide. 
For sample preparation the cover slide was removed and 3$\mu$l of cell culture were 
spread on the agarose pad, which was then recovered with the slide. For each slide, data 
collection did not last for longer than 2h. Data were acquired at room temperature
and none of the bacteria was dividing during a measurement.

\subsection{Optical setup}
Fluorescence correlations spectroscopy (FCS) measurements were performed on a 
LSM Meta 510 system (Carl Zeiss, Jena, Germany) using a 40$\times$ NA 1.2 
UV-VIS-IR C-Apochromat water immersion objective and a home-built detection unit at 
the fiber output channel: A bandpass filter (AHF Analyse Technik, T\"ubingen, Germany) 
was used behind a collimating achromat to reject the residual laser and background 
light. Another achromat (LINOS Photonics, G\"ottingen, Germany) with a shorter focal 
length was used to image the internal pinhole onto the aperture of the fiber of the single
photon counting
avalanche photo diode (SPCM-CD 3017, PerkinElmer, Boston, MA, USA). The correlation curves 
were obtained with a hardware correlator Flex 02-01D (correlator.com, Bridgewater, NJ, 
USA). High magnification laser scanning microscope (LSM) images were taken of 
the area of interest and the detection volume was placed at the desired position, 
along the small dimension always in the center, so that there were mainly hoizontal 
parts of the membrane. Since the LSM and the FCS use the same beam path, the
correspondence between FCS spot and LSM image is excellent. The z-position of the 
spot was stable for many minutes with an accuracy of 100nm, see \cite{chia06} where 
the same experimental setup had been used. No high frequency oscillations in the
image plane were detectable in the alignment correlation curves. We did not observe
any drift within 40 consecutive measurements of 5s each. The waist $w_{0}$ 
of the detection volume was determined 
in calibration measurements with the fluorescent dye Alexa Fluor 488 diffusing freely
in water to be $w_{0}=157\pm12$nm.

\subsection{Theoretical autocorrelation curves}
The experimental autocorrelation curves were analyzed by fitting 
autocorrelation curves expected for different processes. Since the height of the 
detection volume is larger than the diameter of the bacterium, the cytoplasmic 
diffusion can be approximated to occur in two dimensions. Fitting with a more 
refined model taking into account the geometry
of the detection volume in the bacterium~\cite{genn00} did not significantly change the 
values we obtained assuming the simplified geometry. For two independent species 
diffusing with respective diffusion constants $D_{1}$ and $D_{2}$ the correlation curve 
is~\cite{elso74,kric02}
\begin{equation}
\label{eq:doublediff}
G_{\rm diff}(\tau) = \frac{1}{N_{1}+N_{2}}\left\{F\frac{1}{1+\tau/\tau_{1}} + 
(1-F)\frac{1}{1+\tau/\tau_{2}}\right\}\quad.
\end{equation}
Here, the number fraction of particles of one species is given by 
$F=N_{1}/(N_{1}+N_{2})$, where $N_{1}$ and $N_{2}$, respectively, are the average 
numbers of particles of the different species in the detection volume. The characteristic 
relaxation times $\tau_{1}$ and $\tau_{2}$ are linked to the respective diffusion 
constants and the width $w_{0}$ of the detection volume through 
$\tau_{i}=w_{0}^{2}/(4D_{i})$, $i=1,2$.  For particles changing between a mobile state 
(diffusion constant $D$)
and an immobile state, we assume the following 
reaction kinetics for the fraction $F$ of the mobile state $dF/dt = -F/\tau_{1} + 
(1-F)/\tau_{2}$, where $\tau_{1}$ and $\tau_{2}$ are the cytoplasmic and membrane residence
times, respectively. The autocorrelation of the fluctuations has the 
form~\cite{elso74,kric02}
\begin{equation}
\label{eq:exchange}
G_{\rm ex}(\tau)=\frac{(2\pi)^{-2}w_{0}^{2}}{(N_1+N_2)}\ \int_0^{\infty} dk \ k \ 
e^{-\frac{w_{0}^2}{4}(k_x^2+k_y^2)}\left\{ A_{1}e^{\lambda_1 \tau}  +
A_{2}e^{\lambda_2 \tau} \right\}\quad,\nonumber
\end{equation}
where $\lambda_{1,2}=-(Dk^{2}+\tau_{1}^{-1}+\tau_{2}^{-1})/2\pm\left\{
(D k^{2}+\tau_{1}^{-1}+\tau_{2}^{-1})^{2}-4Dk^{2}/\tau_{2}\right\}^{1/2}/2$,
$A_{1,2}=\left\{\lambda_{2,1}+Dk^{2}\tau_{1}/(\tau_{1}+\tau_{2})\right\}/(\lambda_{2,1}-\lambda_{1,2})$. 
For a single species diffusing anomalously in 
two dimensions the autocorrelation function is given by~\cite{schw99a}
\begin{equation}
\label{eq:singleanodiff}
G_{\rm a}(\tau) = \frac{1}{N}\frac{1}{1+\left(\frac{\tau}{\tau_{\rm a}}\right)^{\alpha}}\quad.
\end{equation}
Here, $\tau_{\rm a}^{-\alpha}=4\Gamma/w_{0}^{2}$, where the anomalous exponent 
$\alpha$ governs the spreading of an initially localized distribution 
$\langle x^{2}\rangle\sim t^{\alpha}$ and where $\Gamma$ is the anomalous transport 
coefficient.

Since the cytosplasmic pH of \textit{E.~coli} is about 7.7~\cite{pada76}, 
pH-dependent blinking can be neglected~\cite{haup98}.

\subsection{Data analysis}
The correlation curves were fitted in the time interval 
$\tau\in[2\mu s,1s]$ with a weighted nonlinear least-squares fitting 
algorithm. Curves were selected automatically based on convergence 
of the fit algorithm and goodness of the fit ($\chi^{2}<1.2$ for EGFP and $\chi^2<1.4$ 
for Min proteins). For the Min proteins, curves were first hand-selected for low and
high intensity phases and then selected automatically for quasi-steady states. The latter was 
checked by requiring a constant fluorescence intensity during the measurement. \\

\section{Results and discussion}

\subsection{EGFP}  
\begin{figure}[H]
\includegraphics{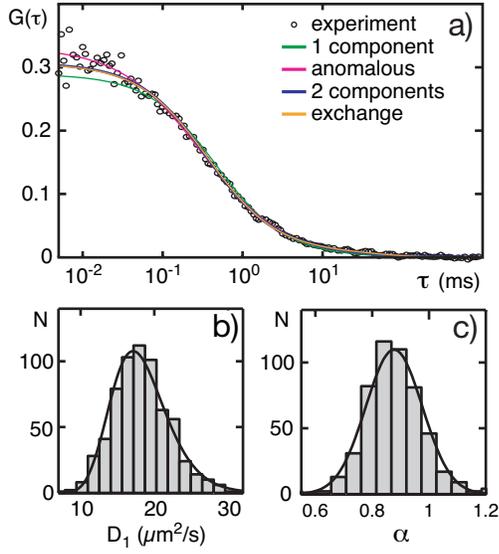}
\caption{\label{fig:EGFP}Diffusion coefficients of EGFP in \textit{E.~coli} measured 
by fluorescence correlation spectroscopy. a) Typical autocorrelation $G(\tau)$ for 
EGFP (black circles) and non-linear least square fits of correlation curves expected 
for different processes. Green: diffusion, see (\ref{eq:doublediff}) with $F=1$, gives
$D=12.9\pm2.3\mu{\rm m}^2/s$ with $\chi^{2}=1.6$. Pink: anomalous diffusion, 
see {eq:singleanodiff}), yields
$\alpha=0.83\pm0.01$ and $\Gamma=4.7\pm0.75{\mu{\rm m}^{\alpha}}/{s}$ with 
$\chi^{2}=1.1$. Blue: two independent diffusing 
populations, see (\ref{eq:doublediff}), yields $D_{1}=17.7\pm3.6\mu{\rm m}^2/s$,
$D_{2}=0.3\pm0.2\mu{\rm m}^2/s$, and $F=0.96\pm0.01$ with
$\chi^{2}=1.1$. Yellow: exchange between a mobile and an immobile state, 
see (\ref{eq:exchange}), yields $D=14.8\pm2.8\mu{\rm m}^2/s$, 
$\tau_{1}=2.3\pm1.0$s, and $F=0.97\pm0.004$ with $\chi^{2}=1.1$
No significant autofluorescence of cells was detected, but there was a non-correlated 
background of 8 kHz from the medium. b) Histogram of 
diffusion coefficients obtained from fitting $G_{\rm diff}$ to 1020 measurements. 
Solid line: log-normal distribution with geometric mean 
$D=17.9^{+4.3}_{-3.4}\mu{\rm m}^2/s$. c) Histogram 
of anomalous exponents from fitting $G_{\rm a}$ to the same curves as in (b). 
Solid line: normal distribution 
with mean $\alpha=0.88$ and variance $\sigma_{\alpha}^{2}=0.09$
In (b) and (c) only fits with $\chi^{2}<1.2$ were considered.}
\end{figure}

We first measured the autocorrelation of the fluorescence fluctuations 
of Enhanced Green Fluorescent Protein (EGFP) in living \textit{E.~coli}, see Materials 
and Methods.  A typical correlation curve is depicted in figure~\ref{fig:EGFP}a. From a 
fit of the correlation curve $G_{\rm diff}$ (\ref{eq:doublediff})  expected for a single diffusing 
species with $F=1$ an apparent diffusion 
constant of $D=12\pm2.3\mu{\rm m}^2/s$ is obtained. 
There are two sources contributing to the error in the value of the diffusion constant. First,
a systematic error results from uncertainties in determining the size of the detection volume.
The size of the detection volume is needed for transforming the relaxation time that can
be extracted from the correlation curve into a diffusion constant.
We estimate this error to be 15\%. Secondly, the fit of the expected correlation curve 
to the data is of finite accuracy due to noise present in the experimental correlation curve 
(around  $10\%$). For the curve in figure~\ref{fig:EGFP}a, the fit quality is reasonable with 
$\chi^{2}=1.58$. In view of the measurements on MinD and MinE, other models were used for 
analyzing the correlation curves. Fitting the data to the autocorrelation $G_{\rm diff}$ (\ref{eq:doublediff})
expected for two independent populations of diffusing particles, where $F$ is 
now a fit parameter, the fit quality was significantly improved, $\chi^{2}=1.08$. For the curve 
in figure~\ref{fig:EGFP}a, the apparent diffusion 
constant of the fast component is $D_{1}=15.6\pm3.2\mu{\rm m}^2/s$.  
Furthermore, we considered the case of molecules switching between a mobile and an 
immobile state. The corresponding autocorrelation is $G_{\rm ex}$, see (\ref{eq:exchange}). 
For the diffusion constant in the 
mobile state, we found $D=14.8\pm5.0\mu{\rm m}^2/s$ with $\chi^{2}=1.08$.
Previous reports suggest deviations 
from normal diffusion of EGFP in vivo, which was attributed to crowding in the cellular 
environment~\cite{weis04}. We therefore considered anomalous diffusion of EGFP, 
where the mean square displacement grows as $\sim t^{\alpha}$ with $\alpha<1$. 
Fitting the correlation $G_{\rm a}$ 
(\ref{eq:singleanodiff}) we obtained an 
anomalous exponent of $\alpha=0.85\pm0.14$ and an anomalous transport coefficient 
$\Gamma=5.9\pm0.94{\mu{\rm m}^{2}}/{s^{\alpha}}$ with $\chi^{2}=1.07$. As can be 
seen in figure~\ref{fig:EGFP}a, the different fits are barely distinguishable.

A histogram of the diffusion constants obtained by fitting $G_{\rm diff}$
to 1021 curves is presented in figure~\ref{fig:EGFP}b.
The histogram is well described by a log-normal distribution with a geometric mean
of $D=17.9^{+4.3}_{-3.4}\mu{\rm m}^2/s$. Within the accuracy of our 
measurements, different cells give the same value for the EGFP diffusion 
constant. 
Hand-selection of curves as is often done in FCS measurements reduced 
the $1\sigma$-confidence interval, but did not change the geometric mean. 
The 
fraction of the fast component was $F=0.96\pm0.03$, indicating that most of the 
dynamics can be attributed to diffusion. We arrived at the same conclusion using 
$G_{\rm ex}$ for the data analysis, see Table~\ref{table:diff}.
Figure~\ref{fig:EGFP}c presents a histogram of anomalous exponents from analyzing 
the same curves using $G_{\rm a}$.  The mean value is $\alpha=0.88\pm0.1$ 

The values of the diffusion constants are surprisingly large in view of previous measurements 
of the EGFP diffusion constant using FRAP, yielding $D_{\rm GFP}\simeq7.5\mu $m$^{2}/s$, 
see~\cite{elow99}. There, it was also found that the diffusion constant can be changed 
significantly by adding a His-tag. We examined His6-EGFP expressed in the same
strain as was used for the measurement of EGFP mobility. Using either
$G_{\rm diff}$ or $G_{\rm ex}$, we found a decrease in the diffusion constant
of about 20\% compared to EGFP. Based on the 
anomalous diffusion model, we found a slightly reduced value for the anomalous
mobility, $\Gamma=5.6^{+5.7}_{-2.8}{\mu{\rm m}^{2}}/{s^{\alpha}}$, while the
anomalous exponent remained the same, $\alpha=0.88\pm0.1$ 
  
\subsection{Quasi-steady states during Min-oscillations} 
\begin{figure}[H]
\includegraphics{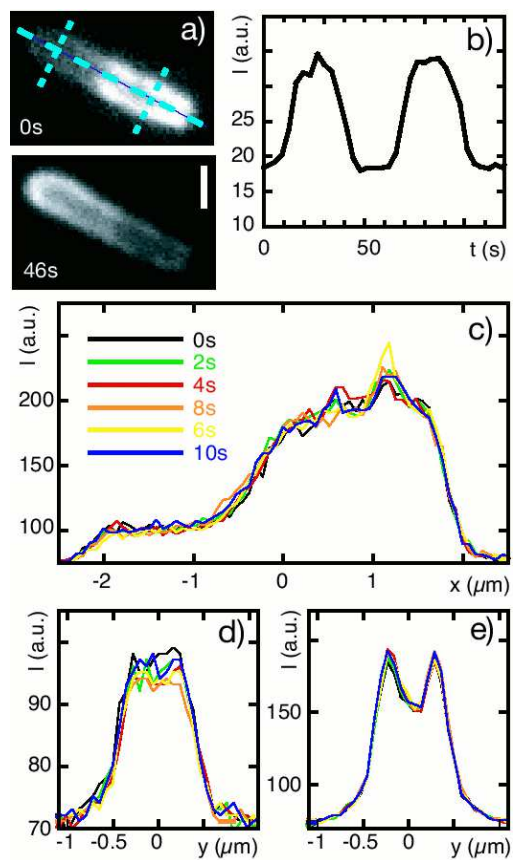}
\caption{\label{fig:qss}Quasi steady states. a) GFP-MinD fluorescence
in \textit{E.~coli} at different phases of the oscillation cycle. Scale bar: $1\mu$m.
b) Fluorescence intensity in a confocal volume located in one cell half as a function 
of time. Oscillations with a period of $~60s$ are clearly seen. Around states of maximal
and minimal intensity, time-intervals of roughly constant fluorescence intensity
can be detected. c,d,e) Fluorescence intensity along the long axis (c) and the 
cross-sections (d, e) indicated in (a) for six different times separated by 2s each. 
The color code for all three panels is as given in (d).
The curves vary around a quasi-stationary mean profile. The differences in the 
cross-section profiles (d) and (e) reflect the different fractions of membrane-bound 
proteins in the low- and high-intensity phases in a cell half, respectively.}
\end{figure}

The analysis of 
fluorescence fluctuations requires a well-defined average state. Seemingly,
this is not the case for the Min-system, which oscillates with a period of about 
80s~\cite{rask99,hu99,meac05}, see figure~\ref{fig:qss}a. However, there are regions 
in the bacterium in which the fluorescence signal is quasi-stationary for about
10s. In figure~\ref{fig:qss}b, we present the fluorescence intensity in a confocal 
volume positioned in one cell half. There are phases
of high and low constant fluorescence as well as phases of strongly varying
fluorescence. Respectively, these phases reflect the dwelling of MinD in one cell 
half for a large
fraction of a half-period as well as the comparatively rapid transition to the opposite
cell half. Figure~\ref{fig:qss}c displays the fluorescence intensity along the bacterial 
long axis for six different times separated by 2s. The intensity variations during this
period are less then $5\%$.  The fluorescence profiles
in cross-sections perpendicular to the long axis also show only moderate fluctuations, 
figure~\ref{fig:qss}d,e. The form of the mean profiles in the low- and high-intensity regions 
differ significantly: while the profile in the low-intensity region is uni-modal, it
is bi-modal in the high-intensity region. This results from a low fraction of 
membrane-bound MinD in the low-intensity region and a high fraction in the 
high-intensity region~\cite{rask99}. 
The fluorescence profiles for different times then indicate that 
the respective amounts of cytoplasmic and membrane-bound MinD are 
quasi-stationary within the 10s shown.

\subsection{GFP-MinD} 
\begin{figure}[H]
\includegraphics{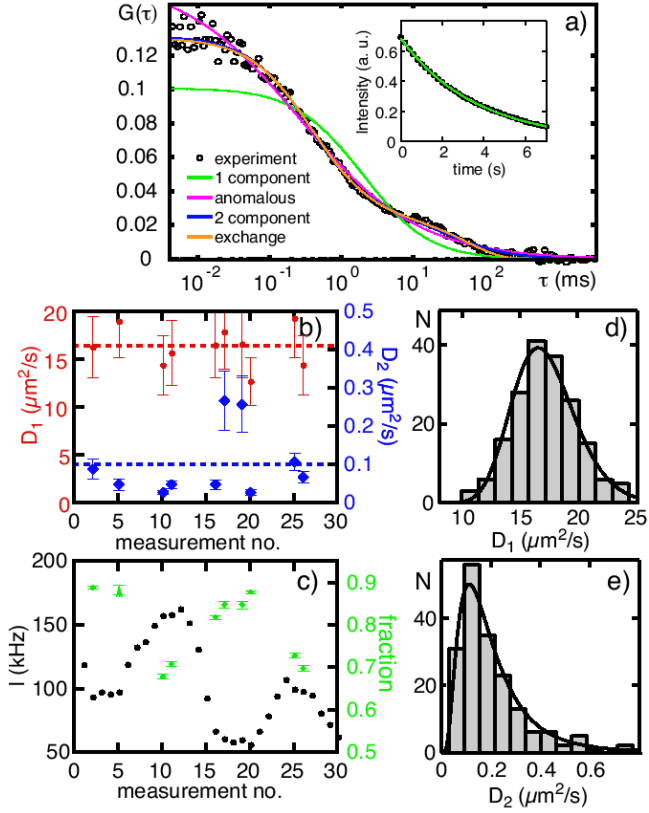}
\caption{\label{fig:MinDAC}Correlation analysis of MinD mobility - two independent
diffusing species. a) Typical autocorrelation curve for GFP-MinD in a region of 
quasi-steady state (black circles) and non-linear least square fits of different expected 
correlation curves. Green and pink: diffusion and anomalous diffusion, respectively. 
Essential features of the experimental curve are missed ($\chi^{2}=5.6$ and $1.8$, respectively).
Blue: two independent diffusing populations, see (\ref{eq:doublediff}), yields 
$D_{1}=19.8\pm4.3\mu{\rm m}^2/s$, $D_{2}=0.11\pm0.02\mu{\rm m}^2/s$,
and $F=0.74\pm0.01$ with $\chi^{2}=1.1$. Yellow: exchange between a diffusing and 
an immobile state yields $D=15.7\pm3.1\mu{\rm m}^2/s$, 
$\tau_{1}=302\pm25$ms, and $F=0.83\pm0.004$ with $\chi^{2}=1.18$.  Inset: 
Typical fluorescence intensity of EGFP adsorbed on a glass slide as a function of time. 
Photobleaching reduces the intensity. The green line is an exponential fit with a decay time of 3.6s. 
The same laser power as for measurements on bacteria was chosen. b) Apparent 
diffusion constants $D_{1}$ and $D_{2}$ for 10 curves admitting a good fit 
($\chi^{2}<1.4$) 
among 30 successive measurements on a single cell. The mean values are 
$D_{1}=16.4\pm2.1\mu{\rm m}^2/s$ (mean$\pm$SD) and 
$D_{2}=0.1\pm0.09$ (mean$\pm$SD). c) Fluorescence 
intensity and
fast fraction for the same measurements as in (b). The fast fraction is higher for low
intensities. Error bars in (b) and (c) indicate the 95\% confidence interval of the fit.
Additional statistical errors can be expected.
d,e) Histograms of the diffusion constants.
Only curves with quasi-steady fluorescence intensity and a fit quality of $\chi^{2}<1.4$
were retained. Solid lines: log-normal distributions with geometric
means $D_{1}=17.0^{+3.0}_{-2.5}\mu{\rm m}^2/s$ and 
$D_{2}=0.17^{+0.14}_{-0.08}\mu{\rm m}^2/s$.}
\end{figure}

We measured MinD-motility  in the strain JS964. For the FCS analysis, we considered only
fluorescence curves taken from regions in quasi-steady state. Every individual 
measurement lasted for 5s. A typical autocorrelation curve is shown in 
figure~\ref{fig:MinDAC}a. From the graph it is obvious that two distinct time-scales are 
present. We first checked that neither of them is due to bleaching. To this end we adsorbed 
EGFP on an untreated cover slip. Then we recorded intensity traces and correlation curves 
for this immobilized EGFP. The intensity curves could be fitted to an exponential curve 
with a decay time of a few seconds, see figure~\ref{fig:MinDAC}a inset. The corresponding 
FCS curves show a decay with a similar characteristic time. These times are larger than the
two time-scales apparent in figure~\ref{fig:MinDAC}a. Furthermore, the correlation curves were
largely independent of the excitation intensity (data not shown). We conclude 
that neither of the time-scales is due to bleaching of immobilized molecules. To reduce the
already weak contribution to the correlations by bleaching of immobilized molecules even further, 
we recorded the first correlation curve in an experiment only a few seconds after the laser
was switched on.

One of the time-scales detectable in figure~\ref{fig:MinDAC}a is readily attributed to MinD 
diffusing freely in the cytoplasm. The existence of MinD bound to the membrane suggests 
two obvious candidate processes leading to the other 
time-scale visible in the correlation curves. First of all, it could 
be attributed to the diffusion of MinD on the membrane. Secondly, it could result from the 
exchange of MinD between the membrane and the cytoplasm. We analyzed the measured 
correlation curves using separately the two different models. Of course, the two processes
are not mutually exclusive. It would thus be desirable to analyze the correlation curves 
using a model that accounts for diffusion on the membrane as well as for binding and unbinding. 
However, the expected correlation curve differs only by small amounts from the curves for either
of the two alternatives separately.  The accuracy of our measurements does not allow distinguishing 
between them. Note, that a significant fraction of membrane-bound MinD might be immobile as
it is incorporated into helices~\cite{shih03}. Since these molecules do not contribute to fluctuations 
in the average fluorescence intensity, FCS cannot detect them. 

We first present the results assuming two states of different mobility. 
Figure~\ref{fig:MinDAC}b displays the two diffusion constants obtained from fits of 
$G_{\rm diff}$ (\ref{eq:doublediff}) to different correlation curves measured on a 
single cell. We interpret the faster diffusion constant to represent the mobility of cytoplasmic 
MinD. It is of the same order as the diffusion constant of EGFP, see Table~\ref{table:diff}. 
The smaller diffusion constant is interpreted as resulting from the mobility of membrane-bound MinD. 
This is supported by the estimated value of the fraction of the fast component. In agreement
with the measurements of the cross-sections, figure~\ref{fig:qss}d, e, the fraction of fast 
moving proteins is larger in the low-intensity regions than in the high-intensity regions, 
see  figure~\ref{fig:MinDAC}c. The difference is 10 to 15\%, less than one might have expected 
from an investigation of the cross-sectional profiles in figure~\ref{fig:qss}d and e. As mentioned
in the previous paragraph,  FCS possibly overestimates the fraction of cytoplasmic proteins because some
fraction of membrane-bound MinD might be immobile as it forms helices. Note, that the standard 
deviation of the mean diffusion constant is smaller than the estimated error of a single measurement, 
showing that the quality of our results is not limited by variations within a cell.

Histograms of fast and slow diffusion constants summarizing series of measurements
on different cells are shown in figure~\ref{fig:MinDAC}d,~e. 
Both histograms are well described by a log-normal distribution. The geometric mean 
value for the fast diffusion constant is $D_{1}= 17.0^{+3.0}_{-2.5}\mu{\rm m}^2/s$. 
For the slow diffusion constant we find 
$D_{2}= 0.17^{+0.14}_{- 0.08}\mu{\rm m}^2/s$. This value is one order of 
magnitude higher than the diffusion constant for the transmembrane histidine 
kinase PleC measured by single protein tracking in \textit{C.~crescentus}~\cite{deic04}. 
Since PleC is a transmembrane protein, while MinD binds to the 
polar heads of the lipids forming the membrane, the values seem to be compatible. No 
correlation could be detected between the values of the fast and slow diffusion constants 
(data not shown). Separating the curves into those with low and high average intensity 
does not reveal significant differences between the respective fast and slow diffusion 
constants, see Table~\ref{table:diff}. In the low-intensity regions, however, the fraction 
$F=0.81\pm0.1$ of the fast-diffusing component is larger  than in the high-intensity 
regions, where $F=0.71\pm0.1$. The difference in the fractions is more pronounced 
when averaging over several measurements on a single cell than when averaging over
measurements on different cells, figure~\ref{fig:MinDAC}c. This presumably reflects 
different protein concentrations in different cells.

\begin{figure}[H]
\includegraphics{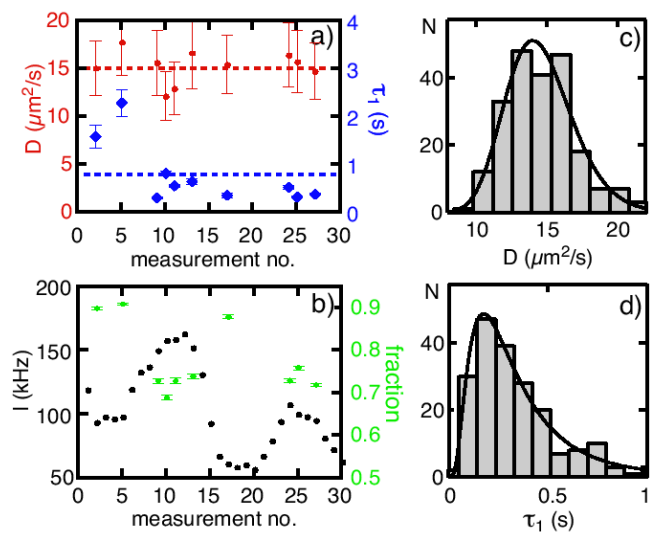}
\caption{\label{fig:MinDCA}Correlation analysis of MinD mobility - exchange between
diffusing and immobile state. a) Apparent diffusion constants and residence times in the 
mobile state for the same 30 successive measurements on a single cell as in 
figure~\ref{fig:MinDAC}b,c. The mean values are $ D=15.0\pm1.9\mu{\rm m}^2/s$  
and $\tau_{1}=783\pm651$ms (mean$\pm$SD). b) Fluorescence instensity and 
mobile fraction for the same measurements as in (a). The mobile fraction is higher for 
low intensities. Error bars in (a) and (b) indicate the 95\% confidence interval of the fit.
 c,d) Histograms of the diffusion constants and residence times obtained 
from the same 2017 measurements as in figure~\ref{fig:MinDAC}d,e. Solid lines:
log-normal distributions with geometric means 
$D=14.4^{+2.6}_{-2.2}\mu{\rm m}^2/s$ and $\tau_{1} =322^{+422}_{-183}$ms.}
\end{figure}

We analyzed the same data based on the exchange of MinD between a mobile
(cytoplasmic) state and an immobile (membrane-bound) state, disregarding diffusion of 
membrane-bound proteins.  As suggested by the 
cross-section profiles, figure~\ref{fig:qss}d, e,  we assume the average fraction of mobile 
molecules to be constant during one measurement. In that case, the residence times 
$\tau_{1}$ and $\tau_{2}$ of MinD in the mobile and immobile states, respectively, are 
related to the fraction $F$ of mobile molecules by $F=\tau_{1}/(\tau_{1}+\tau_{2})$. The 
results obtained from analyzing the same curves as in figure~\ref{fig:MinDAC}b, c are 
displayed in figure~\ref{fig:MinDCA}a, b. The diffusion constants are in the same range as 
the values of the fast diffusion constant obtained above. The same holds for the value of 
the mobile fraction $F$. Histograms of the diffusion constant and the residence time in the 
mobile state are presented in figure~\ref{fig:MinDCA}c, d. Differences in the values for 
low- and high-intensity regions are not significant, although the residence times are on 
average larger in the low-intensity regions, see Table~\ref{table:diff}. We repeated the 
measurement using a different strain (WM1255). The average cytoplasmic diffusion
constants are smaller in this strain, while the average residence time is a little larger, see 
Table~\ref{table:diff}. In view of the broadness of the distributions, however, the differences 
are not significant.

\subsection{MinE-GFP} 
\begin{figure}[H]
\includegraphics{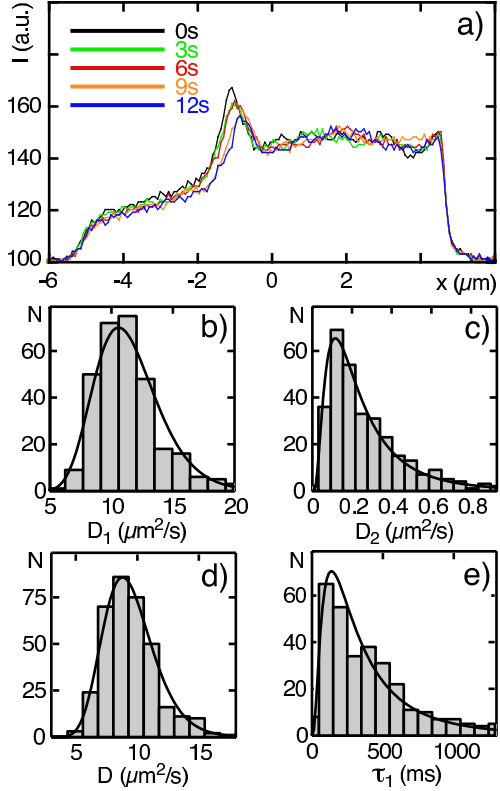}
\caption{\label{fig:MinE}Correlation analysis of MinE. a) Quasi-steady state of the 
MinE distribution along a cell's long axis. Five curves separated each by 3s vary
around a mean profile. An accumulation of MinE close to the cell center, commonly
known as MinE ring, can clearly be recognized. It moves slowly to one cell pole. 
b,c) Histograms of the diffusion constants assuming two independent diffusing species. 
Only curves with quasi-steady fluorescence intensity and a fit quality of $\chi^{2}<1.4$ 
were retained. Solid lines: log-normal distributions with geometric means 
$D_{1}=11.2^{+2.9}_{-2.3}\mu{\rm m}^2/s$ and 
$D_{2}=0.20^{+0.23}_{-0.11}\mu{\rm m}^2/s$. d,e) Histograms of the diffusion constants and residence times obtained from the same  measurements as in (b,c) 
assuming exchange between a diffusing and an immobile state. Solid lines: log-normal 
distributions with geometric means $D=9.3^{+2.3}_{-1.9}\mu{\rm m}^2/s$ and 
$\tau_{1} =396^{+888}_{-274}$ms.}
\end{figure}

For measuring the mobility of MinE we employed the same strategy 
as for MinD. An example of a quasi-steady state of the MinE distribution is shown in 
figure~\ref{fig:MinE}a. As for MinD, two distinct relaxation times can be detected in the 
correlation curves. We analyzed these curves using the same models as for MinD. 
Histograms of the two different diffusion constants and of the diffusion constant together 
with the residence time in the mobile state, respectively, are presented in 
figure~\ref{fig:MinE}b-e. As before, the histograms are well described by log-normal 
distributions. Assuming two independent populations with different mobilities, we find 
$D_{1}=11.2^{+2.9}_{-2.3}\mu{\rm m}^2/s$ and 
$D_{2}=0.20^{+0.23}_{-0.11}\mu{\rm m}^2/s$. The fraction of the faster diffusion 
population is $F=0.79\pm0.10$. While cytoplasmic diffusion of MinE is thus
smaller than of MinD, the diffusion constants for membrane-bound MinD and MinE are
the same. This is compatible with MinE being bound to MinD on the membrane.
Assuming the other model, we obtain for MinE
$D=9.3^{+2.3}_{-1.9}\mu{\rm m}^2/s$ and $\tau_{1}=396^{+888}_{-274}$ms.
The mobile fraction is in this case $F=0.86\pm0.09$. Separating the curves into those
from a low-intensity and those of a high-intensity phase, no significant differences
between neither the values of the diffusion constants nor the residence times in the 
different phases can be detected, see Table~\ref{table:diff}.

\section{Conclusion and outlook}

In the present work we have used FCS to determine Min-protein mobility in living
\textit{E.~coli}. The possibility to apply FCS relies on the existence
of quasi-stationary steady states in some regions of the bacterium for time
intervals of at least 10s, see figure~\ref{fig:qss}c-e and \ref{fig:MinE}a. Our 
correlation data clearly show the existence of more than one 
relaxation time, which can satisfactorily be explained by assuming for both
MinD and MinE two states of different mobility. We interpret the faster component as
resulting from diffusion of cytoplasmic proteins. The second time-scale could result 
from the mobility of proteins in the membrane-bound state or from transitions between 
the cytoplasm and the membrane. We find that all in all both models fit equally well to the 
data, even though for individual curves there can be significant differences in the fit quality. 
Using either of the corresponding correlation curves, $G_{\rm diff}$ or $G_{\rm ex}$, for analyzing
the experimental data, we find values around 16$\mu$m$^{2}/s$ and 10$\mu$m$^{2}/s$
for the respective cytoplasmic diffusion constants of GFP-MinD and MinE-GFP. Therefore,
a cytoplasmic MinD molecule explores the volume of a 4$\mu$m long cell within
roughly a second. cytoplasmic MinE, which readily forms dimers, needs about 1.5s, i.e., only 
slightly longer.

The diffusion constants we measured for membrane-bound
proteins are about two orders of magnitude smaller than the cytoplasmic diffusion constants.
For membrane-bound MinD, it is of the same order as the value assumed in the AC model 
studied in \cite{meac05}. This shows that the mobility of membrane-bound MinD
is sufficiently large to allow for an AC mechanism causing the oscillations. In CA models it is
usually assumed that membrane-bound proteins are immobile. However, it was found in 
\cite{fang06} that oscillations can still be generated in a CA model if diffusion of MinD 
on the membrane is two orders of magnitude smaller than in the cytoplasm.

For the average residence time of MinD in the 
cytoplasm we find a value of about 300ms. In order to generate ``striped'' patterns
in long bacteria, the CA model introduced in \cite{huan03} requires the exchange
of ATP for ADP on cytoplasmic MinD to be not too fast. For the parameters used there, the 
authors find an upper critical rate of 1/s. Since the measured residence time provides a 
lower bound on the exchange rate of about 3/s (only after rebinding of ATP can MinD 
attach again to the membrane), a re-investigation of the model is in order. The residence
time of MinE in the cytoplasm is somewhat larger than for MinD which is compatible with 
the fact that MinE requires MinD as a substrate in order to bind to the membrane.
From the residence time in the cytoplasm and the cytoplasmic diffusion constants, we can 
determine the diffusion length $\ell=(Dt)^{1/2}$. This is the average distance travelled
by a cytoscolic molecule. For MinD and MinE we find $\ell\simeq2\mu$m. This value indicates
that in small bacteria of about 2$\mu$m in length, the distribution of cytoplasmic MinD and
MinE should be homogenous. As AC models, but not CA models produce oscillations 
under these conditions, a detailed investigation of short cells might be helpful. Particular
attention should be paid to the MinE-ring in these cells. The reason is that the investigation
of the CA model by Huang et al.~\cite{huan03} suggests disappearance of the MinE-ring 
if $\ell$ is increased in comparison to the cell length. The presence or absence of the
MinE-ring in short cells might therefore provide interesting information on the mechanism of its
formation.

Comparing the different values measured in high- and low-intensity phases, respectively,
we find that the fraction of cytoplasmic proteins is always larger in the low-intensity phases.
This is not only an effect due to averaging but is also present in individual cells, see 
Figs.~\ref{fig:MinDAC}c and \ref{fig:MinDCA}b.
Based on the CA models, a shorter cytoplasmic residence time of MinD
in the high-intensity phase than in the low-intensity phase is expected. Indeed, 
on average, our measurements confirm this expectation, see Table~\ref{table:diff}. 
Caution should be taken, though, because the error bars are quite large. The 
average residence time of MinE in the cytoplasm, too, depends on being in a high- 
or low-intensity phase. This is expected since a higher number of membrane-bound 
MinD should lead to a higher rate of MinE binding to the membrane. 

The results presented here are compatible with the mechanism
of forming MinD aggregates on the membrane suggested in~\cite{zhou04}: cytoplasmic MinD
dimerizes. As a consequence the membrane targeting sequence which is associated with
the MinD's C-teminal helix is exposed and the dimer binds to the membrane. Subsequently,
MinD-aggregates are formed through attractive interactions between the membrane-bound
dimers. In order to test this hypothesis further,
mobility measurements on mutant proteins might be helpful. Furthermore, the consequences
of this mechanism for the Min-oscillations have to be explored by theoretical analysis.  

\ack We thank W. Margolin for donation of the strains
WM1079 and WM1255 and J. Lutkenhaus for donation of the strain JS964 and 
R. Hartmann for help with the expression of EGFP and preparation of bacteria.

\section*{Glossary}

\begin{description}
\item[Cytoplasm.] The internal content of a cell (in eukaryotes: except the nucleus). It is surrounded 
by the cytoplasmic membrane, a lipid bilayer.
\item[Fluorescence Corrleation Spectroscopy (FCS).] A spectroscopy method that exploits
fluorescence fluctuations around an average value. Fluorescence fluctuations can be due
to mobility of the fluorophore or due to chemical reactions.
\item[Green Fluorescent Protein (GFP).] A protein from the jellyfish \textit{Aequorea
aequorea} that fluoresces green with an emission maximum at 509nm when exposed to blue light. 
EGFP is a bright mutant of GFP with an emission maximum at 511nm. The molecular weight
is about 27kD.
\item[Min system.] A set of proteins involved in the determination of the
division site in bacteria. Mutations in these proteins lead to the formation of
not viable small cells (mini-cells).
\item[MinD.] Protein of the Min system with a molecular weight of about 30kD. It is able to hydrolyse
ATP. MinD-ATP has a high affinity for the cytoplasmic membrane, while MinD-ADP is cytoplasmic.
\item[MinE.] Protein of the Min system in \textit{E. coli} with a molecular weight of about 10kD.
When bound to MinD, it is able to speed up ATP-hydrolysis by MinD.

\end{description}

\fulltable{\label{table:diff}
Mobility of 
EGFP, His6-EGFP, GFP-MinD, MinE-GFP. For the Min proteins, curves from 
low-intensity phases (l.i.) and high-intensity (h.i.) phases were analyzed separately.
$N_{\rm tot}$: total number of correlation curves analyzed.
$D_{1}$, $D_{2}$: diffusion constants for two independent populations, $D$, 
$\tau_{1}$: diffusion constant and residence time in the mobile state for proteins
switching between a mobile and an immobile state, $F$: fraction of the faster/mobile
population, $N$: number of curves allowing for a sufficiently good fit. Values were 
considered only from curves where the fit produced a $\chi^{2}<1.4$ (for EGFP 
$\chi^{2}<1.2$) and where the intensity was constant. Displayed are 
the mean values and the $1\sigma$ confidence interval. For EGFP, the values of 
$D_{1}$ and $D$ are well described by a log-normal distribution. The values 
of $D_{2}$ and $\tau_{1}$ scatter extremely and are described neither by log-normal 
nor by normal distributions. 
For the Min proteins, the values of $D_{1}$, $D_{2}$, $D$, and $\tau_{1}$ are 
well described by a log-normal distribution. For all strains, the values of $F$ follow
a normal distribution. $^{a}$BL21(DE3)pLys, $^{b}$JS964, $^{c}$WM1255, 
$^{d}$WM1079.}
\br
& & \multicolumn{4}{c}{two species} & \multicolumn{4}{c}{binding}\\
& $N_{\rm tot}$ & $N_{\rm sel}$&$D_{1}$ ($\frac{\mu{\rm m}^{2}}{s}$) & 
$D_{2}$  ($\frac{\mu{\rm m}^{2}}{s}$) & $F$ & $N$ & 
$D$ ($\frac{\mu{\rm m}^{2}}{s}$) & $\tau_{1}$ (ms) & $F$ & $N$\\
\mr
EGFP$^{a}$& 1021 & &
$17.9^{+4.3}_{-3.4}$ & $0.22^{+0.51}_{-0.16}$ & $0.96^{+0.03}_{-0.03}$ & 652 &
$17.9^{+4.4}_{-3.6}$ & $1100^{+7150}_{-953}$ & $0.97^{+0.04}_{-0.04}$  & 690\\
His6-EGFP$^{a}$ & 555 & &
$14.9^{+3.7}_{-3.0}$ & $0.14^{+0.53}_{-0.11}$ & $0.96^{+0.04}_{-0.04}$ & 214 &
$15.0^{+5.7}_{-4.1}$ & $1870^{+12200}_{-1620}$ & $0.97^{+0.05}_{-0.05}$ & 220\\
\mr
GFP-MinD$^{b}$& 2017 & 438 &
$17.0^{+3.0}_{-2.5}$ & $0.17^{+0.14}_{-0.08}$ & $0.77^{+0.11}_{-0.11}$ & 181 &
$14.4^{+2.6}_{-2.2}$ & $322^{+422}_{-183}$ & $0.79^{+0.11}_{-0.11}$ & 217\\
GFP-MinD$^{b}$ l.i.& & 191 & 
$16.7^{+3.1}_{-2.6}$ & $0.18^{+0.16}_{-0.08}$ & $0.81^{+0.10}_{-0.10}$ & 105 &
$14.7^{+3.0}_{-2.5}$ & $464^{+643}_{-270}$ & $0.86^{+0.08}_{-0.08}$ & 104 \\
GFP-MinD$^{b}$ h.i.& & 247 &
$17.4^{+2.6}_{-2.3}$ & $0.15^{+0.11}_{-0.06}$ & $0.71^{+0.10}_{-0.10}$ & 76 &
$14.1^{+2.2}_{-1.9}$ & $230^{+209}_{-110}$ & $0.73^{+0.10}_{-0.10}$ & 113 \\
\mr
GFP-MinD$^{c}$ & 738 & 102 &
$14.3^{+2.9}_{-2.4}$ & $0.16^{+0.18}_{-0.08}$ & $0.80^{+0.08}_{-0.08}$ & 50 &
$12.4^{+1.8}_{-1.6}$ & $522^{+721}_{-303}$ & $0.84^{+0.07}_{-0.07}$ & 43 \\
\mr
MinE-GFP$^{d}$ & 1807 & 528 &
$11.2^{+2.9}_{-2.3}$ & $0.20^{+0.23}_{-0.11}$ & $0.79^{+0.10}_{-0.10}$& 307 &
$9.3^{+2.3}_{-1.9}$ & $396^{+888}_{-274}$ & $0.86^{-0.09}_{+0.09}$ & 350\\ 
MinE-GFP$^{d}$ l.i.& & 310 &
$11.4^{+2.8}_{-2.3}$ & $0.21^{+0.25}_{-0.11}$ & $0.82^{+0.09}_{-0.09}$& 198 &
$9.6^{+2.5}_{-2.0}$ & $478^{+1105}_{-334}$ & $0.88^{-0.08}_{+0.08}$ & 223\\ 
MinE-GFP$^{d}$ h.i.& & 218 &
$10.9^{+3.1}_{-2.4}$ & $0.20^{+0.20}_{-0.10}$ & $0.75^{+0.11}_{-0.11}$& 109 &
$8.8^{+1.9}_{-1.5}$ & $285^{+542}_{-187}$ & $0.81^{-0.09}_{+0.09}$ & 127\\
\br
\endfulltable

\end{document}